\documentclass[conference]{IEEEtran}
\usepackage{graphicx}
\usepackage{epsfig}
\ifCLASSINFOpdf
\else
\fi
\usepackage[cmex10]{amsmath}
\usepackage{algorithm}
\usepackage{algorithmic}

\usepackage{mdwmath}
\begin{document}
%
\title{Is Light-Tree Structure Optimal for Multicast Routing in Sparse Light Splitting WDM Networks?
}


\author{\IEEEauthorblockN{Fen Zhou, and Mikl\a'os Moln\a'ar}
\IEEEauthorblockA{IRISA / INSA Rennes\\ Campus de Beaulieu\\
Rennes, France, 35042\\
Email: \{fen.zhou, molnar\}@irisa.fr}
\and
\IEEEauthorblockN{Bernard Cousin}
\IEEEauthorblockA{IRISA / University of Rennes 1\\ Campus de Beaulieu\\
Rennes, France, 35042\\
Email: bernard.cousin@irisa.fr}
}

\maketitle

\begin{abstract}
To minimize the number of wavelengths required by a multicast session in sparse light splitting Wavelength Division Multiplexing (WDM) networks, a light-hierarchy structure, which occupies the same wavelength on all links, is proposed to span as many destinations as possible. Different from a light-tree, a light-hierarchy accepts cycles, which are used to traverse crosswise a 4-degree (or above) multicast incapable (MI) node twice (or above) and switch two light signals on the same wavelengths to two destinations in the same multicast session. In this paper, firstly, a Graph Renewal and Distance Priority Light-tree algorithm (GRDP-LT) is introduced to improve the quality of light-trees built for a multicast request. Then, it is extended to compute light-hierarchies. Obtained numerical results demonstrate the GRDP-LT light-trees can achieve a much lower links stress, better wavelength channel cost, and smaller average end-to-end delay as well as diameter than the currently most efficient algorithm. Furthermore, compared to light-trees, the performance in terms of link stress and network throughput is greatly improved again by employing the light-hierarchy, while consuming the same amount of wavelength channel cost.\\
\end{abstract}


%
\IEEEpeerreviewmaketitle

\section{Introduction}
\label{introdcution}
With the inherent capacity to provide high bandwidth and small delay, all-optical Wavelength Division Multiplexing (WDM) networks enable the growth of bandwidth-driven and time sensitive multimedia applications, such as video distribution, multimedia conferencing, and so on~\cite{xdHu2004}. Multicast, which aims to distribute messages simultaneously from the same source to various group members, is highly required to satisfy these applications. Multicast is bandwidth-efficient because it eliminates the necessity for the source to send an individual copy of the message to each destination, and it avoids flooding the whole network by broadcasting~\cite{jyH2002}. However, it is a challenging work to implement multicast in Wide Area Networks (WANs) WDM networks due to high complexity of multicast routing~\cite{xdHu2004}, let alone in spare light splitting~\cite{rMalli1998} WDM mesh Networks, where some nodes namely multicast capable nodes (MC ~\cite{rMalli1998}) can support multicast and the others namely multicast incapable nodes (MI ~\cite{rMalli1998}) cannot. MC nodes are equipped with Splitter-and-Delivery cross-connect~\cite{mAli2000} while MI nodes are equipped with Tap-and-Continue (TaC~\cite{mAli2000Cost}) cross-connect, which is only able to tap into a small amount of light power and forward the rest to one outgoing port. In sparse light splitting WDM networks, multicast routing is to find a set of light distribution structures to cover all the multicast group members under optical constraints. In the absence of wavelength conversion, the same wavelength should be retained over all the links of a light distribution structure.

The main objective of multicast routing and wavelength assignment (MRWA)~\cite{aHamad2006} problem is to optimize the optical network resources in terms of total cost (wavelength channel cost), link stress (maximum number of wavelengths required per fiber), optical power attenuation (impacted by the average end-to-end delay and diameter of the tree) as well as the network throughput. Normally, the light-tree structure~\cite{lhSahasrabuddhe1999} is thought to be optimal and a set of light-trees (or a light-forest~\cite{xjzhang2000}) is employed to accommodate a multicast session. Accordingly, numerous light-trees construction algorithms have been developed such as Reroute-to-Source, Member-First and Member-Only~\cite{xjzhang2000}. Reroute-to-Source makes use of the shortest path tree and hence is optimal in delay and diameter, but its cost and link stress are too big to stand~\cite{fZhou2008Photonic}. Member-Only is based on the Minimum Path Heuristic~\cite{hTakahashi1980} and thus currently through to achieve the best cost and link stress~\cite{aHamad2006,fZhou2008Photonic,xjzhang2000}.

In the case of full light splitting, one light-tree is enough to cover all the multicast members and thus the light-tree structure is optimal of total cost and link stress. But, is the light-tree structure still optimal for sparse light splitting WDM networks? The answer is no. Under splitting constraint, several light-trees may be required to establish one multicast group. The quality of optimization not only depends on the quality of each light-tree but also depends on the number of light-trees built for a multicast session. Given a multicast session, more destinations a light distribution can span, the fewer light distribution structures a multicast session will require. Based on this basic idea, in our study we propose a new multicast structure: light-hierarchy to span as many destinations as possible aiming at improving the link stress and network throughput. Similar to a light-tree, only one wavelength is occupied over all the links in a light-hierarchy; while different from a light-tree, a light-hierarchy accepts cycles. The cycles in a light-hierarchy permit to traverse an at least 4-degree MI node twice (or more) and thus crosswise switch two signals on the same wavelengths to two destinations in the same group by using two different input and output ports pairs. In this paper, a Graph Renewal Strategy is proposed to improve the link stress and total cost of light-trees, and an In Tree Distance Priority is applied to improve the delay and diameter of light-trees. Then, the Graph Renewal Strategy is extended to compute light-hierarchies to improve the multicast performance again in terms of link stress and network throughput.

The rest of the paper is organized as follows. Firstly, the all-optical multicast routing problem is described and the famous Member-Only algorithm is reviewed in Section~\ref{sec: All-Optical Multicast Routing Problem}. Then Graph Renewal Strategy, In Tree Distance Priority and a new multicast structure: light-hierarchy are proposed. Based on these strategies, two multicast routing algorithms, namely GRDP Light-Tree algorithm and GRDP Light-Hierarchy algorithm, are presented in Section~\ref{sec: Proposed Solutions}. Accompanied with the routing problem, the wavelength assignment problem is solved in Section~\ref{sec: Wavelength Assignment}. Numerical results are obtained in Section~\ref{sec: Performance Evaluation And Simulation}. Finally, we conclude this paper in Section~\ref{sec: Conclusion}.

\section{All-Optical Multicast Routing Problem}
\label{sec: All-Optical Multicast Routing Problem}
\subsection{Problem Description}
\label{subsec: Problem Description}
An all-optical WDM mesh network is considered, where light splitters are very sparse and the costly wavelength converters are not available. And we assume that the same wavelength can only be used once in one optical fiber, either in the forward or in the backward direction. A multicast session $ms(s, D)$, where $s$ denotes the source node and $D$ is the set of destinations in the multicast session, is assumed to be required. In order to accommodate this multicast group, a light distribution structure under optical constraints (i.e., wavelength continuity, distinct wavelength~\cite{bMukherjee2000}, sparse light splitting~\cite{rMalli1998} and lack of wavelength conversion) should be built to optimize the network resources such as total cost (i.e., wavelength channels cost) and the link stress (i.e., maximum number of wavelengths required per link). Furthermore, considering the QoS for the time sensitive multimedia applications, the average end-to-end delay needs to be minimized. Taking account of the signal attenuation over distance and the number of amplifiers needed, the diameter (or the height) of the light distribution structures should not be too large. And from the point of view of the network throughput, the call blocking probability (or the inverse of the number of sessions accepted) should be as small as possible. However, not all these parameters could be optimized simultaneously. Here, we are focused on reducing the link stress and improving the network throughput.

\subsection{Previous Work}
\label{subsec: Previous Work}
To facilitate the comparison in the next section, here we review the Member-Only algorithm~\cite{xjzhang2000} deriving from the Minimum Path Heuristic~\cite{hTakahashi1980}. It involves three nodes sets during the construction of the light-trees. In a subtree $LT$ under construction, we maintain the following sets of nodes.

   $MC\_SET$: includes source node, MC nodes and the leaf MI nodes. They may be used to span light-tree $LT$ and thus are also called connector nodes in $LT$.

   $MI\_SET$: includes only the non-leaf MI nodes, whose splitting capability is exhausted. Hence, these nodes are not able to connect a new destination to the subtree $LT$.

   $D$: includes unserved multicast members which are neither joined to the current light-tree $LT$ nor to the previously constructed multicast light-trees.\\

The span of a distribution light-tree $LT$ begins with the source: $LT=\{s\}$, $MI\_SET=\O$, $MC\_SET=\{s\}$ and $D=\{all~destinations\}$. At each step, try to find the nearest destination from $d \in D$ to $c \in MC\_SET$, whose shortest path $SP(d, c)$ does not involve any node in $MI\_SET$. If it is found, $SP(d, c)$ is added to $LT$ and the sets are updated: $d$ and the MC nodes are added to $MC\_SET$, non-leaf MI nodes are added to $MI\_SET$, and remove $d$ from $D$. Otherwise (i.e., no such constraints-satisfying shortest path could be found), the current light-tree $LT$ is finished, and another light-tree assigned a new available wavelength is started using the same procedure until no destination is left in $D$.

\section{Proposed Solutions}
\label{sec: Proposed Solutions}
\subsection{Graph Renewal Strategy}
\label{subsec: Graph Renewal Strategy}
According to the Member-Only algorithm, during the construction of a light-tree, non-leaf MI nodes in the subtree $LT$ (i.e., the nodes in $MI\_SET$) have exhausted their TaC capability, and thus could not be used again to connect another destination to the subtree $LT$. Since they are useless for the spanning of the current light-tree, why don't we delete them from the graph? At each step, in a new graph, say $G_{i}$ (generated by deleting all the non-leaf MI nodes in $LT$ from the original graph $G$), we compute the shortest paths and the distances from the destinations in set $D$ to $LT$. Then, add the nearest destination to $LT$ with the shortest path in $G_{i}$. Here, we can see, it is definitely true that the shortest paths between any two nodes in the new graph $G_{i}$ will not traverse any nodes in $MI\_SET$. Hence, by computing the shortest path in the new graph, when finding the nearest destination to the subtree $LT$, we do not need to check whether its shortest path to $LT$ (precisely speaking, to its connector node in $MC\_SET$ for $LT$) satisfies the light splitting constraint or not.

The Graph Renewal Strategy has two benefits compared to Member-Only algorithm. Firstly, the possible shortest path to connect a destination to a light-tree could be definitely computed out if it exists. Secondly, in case that no constraint-satisfied shortest path could be found, a longer but the shortest one among the constraint-satisfied paths is used to connect a destination to the current light-tree. But with the Member-Only algorithm, only the shortest path is used to span the light-tree and not all the possible shortest path could be enumerated for each node pair. Hence, more available paths could be found to join a destination to the current light-tree with the Graph Renewal strategy. We use the following example to explain the procedure of the Graph Renewal strategy.

\begin{figure}
        \begin{center}
        \includegraphics[width=.4\textwidth]{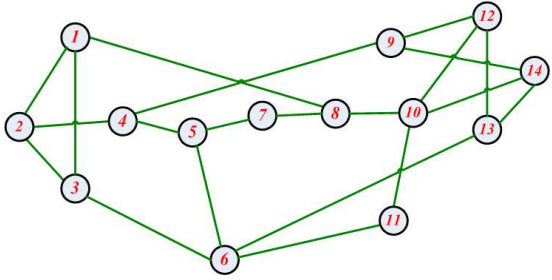}
        \end{center}
        \caption{NSF Network Topology}
        \label{fig: NSFnet}       
    \end{figure}

\begin{figure}
        \centering
        \includegraphics[width=.2\textwidth]{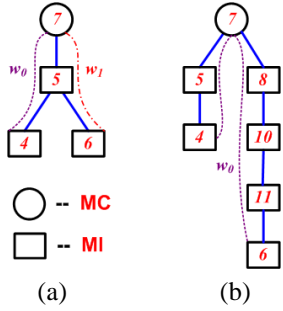}\\
        \mbox{(a)}\hspace{.8in}\mbox{(b)}
        \caption{For multicast session $ms_{1}$, (a) Light-tree built by Member-Only; (b) Light-tree built using Hypo-Steiner Heuristic.}
        \label{fig: hypo}
    \end{figure}

\textbf{Example 1:}
In the NSF network of Fig.~\ref{fig: NSFnet}, a multicast session $ms_{1}\big(s:7,D:(4, 6)\big)$ request arrives, and only the source is an MC node. Using Member-Only algorithm, node 4 is firstly connected to node 7 using the shortest path $SP(7-5-4)$. Now, $MC\_SET = \{4, 7\}$, $MI\_SET = \{5\}$. Next, compute the shortest paths from node 6 to the nodes in $MC\_SET$.  Both $SP(4-5-6)$ and $SP(7-5-6)$ involve non-leaf MI node 5, thus the span of the first light-tree $LT_{1}$ should be stopped and a new light-tree $LT_{2}$ on wavelength $w_{1}$ is required to accommodate node 6 as shown in Fig.~\ref{fig: hypo}(a). But, here if we perform the Graph Renewal Strategy (delete non-leaf MI node 5 in $LT_{1}$ from the original graph $G_{1}$ to get a new graph $G_{2}$), the shortest path $SP_{G_{2}}(7-8-10-11-6)$ in the new graph $G_{2}$ could be found to connect node 6. It is worth noting that $SP_{G_{2}}(7-8-10-11-6)$ is not the shortest path in the original graph, but it is the constraint-satisfied path with the smallest length. As demonstrated in Fig.~\ref{fig: hypo}(b), one light-tree is sufficient to cover all the multicast members, and thus only wavelength $w_{0}$ is required for $ms_{1}=\big(s:7,D:(4, 6)\big)$.

\subsection{In Tree Distance Priority to Improve Delay and Diameter}
\label{subsec: Distance Priority to Improve Delay and Diameter}
The Distance Priority proposed in~\cite{fZhou2008ICCS} could be applied here to reduce the delay and diameter of light distribution structures. The nodes in $MC\_SET$ are assigned priorities according to their distances to the source in the subtree $LT$ (that is why it is called In Tree Distance Priority). Hence, the source itself is associated with the highest priority. This priority is applied when a destination to be added is equally away from more than one connector node in $MC\_SET$. From the point of view of the end-to-end delay and the diameter of light-trees, it is better to add a destination to $LT$ via the connector node nearest to the source in $LT$.

\begin{figure}
        \centering
        \includegraphics[width=.1\textwidth]{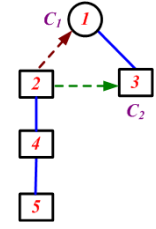}
        \caption{Distance Priority}
        \label{fig: priority}
\end{figure}

\textbf{Example 2:}
Multicast session $ms_{2}\big(s:1,D:(2\sim5)\big)$ request arrives at node 1 in the NSF network in Fig.~\ref{fig: NSFnet}. Seen from the Fig.~\ref{fig: priority}, after the source nodes 1 and 3 are added to the subtree $LT$, node 2 can be connected via both connector nodes 1 and 3.  Since node 1 has higher priority, node 2 is connected via it to the subtree. Then, nodes 4 and 5 are joined. With the In Tree Distance Priority, delay from source node 1 to node 2 is reduced by 1 hop (compared with connected to node 3), and the diameter of the tree is reduced by one hop. Furthermore, the delays from source to those nodes (i.e., nodes 4 and 5) which are joined to the light-tree via node 2 are also reduced. Accordingly, the average end-to-end delay will be reduced too.

\subsection{A New Structure: Light-Hierarchy}
\label{subsec:Edge Disjoint Path Routing Algorithm}
\begin{figure}
        \centering
        \includegraphics[width=.4\textwidth]{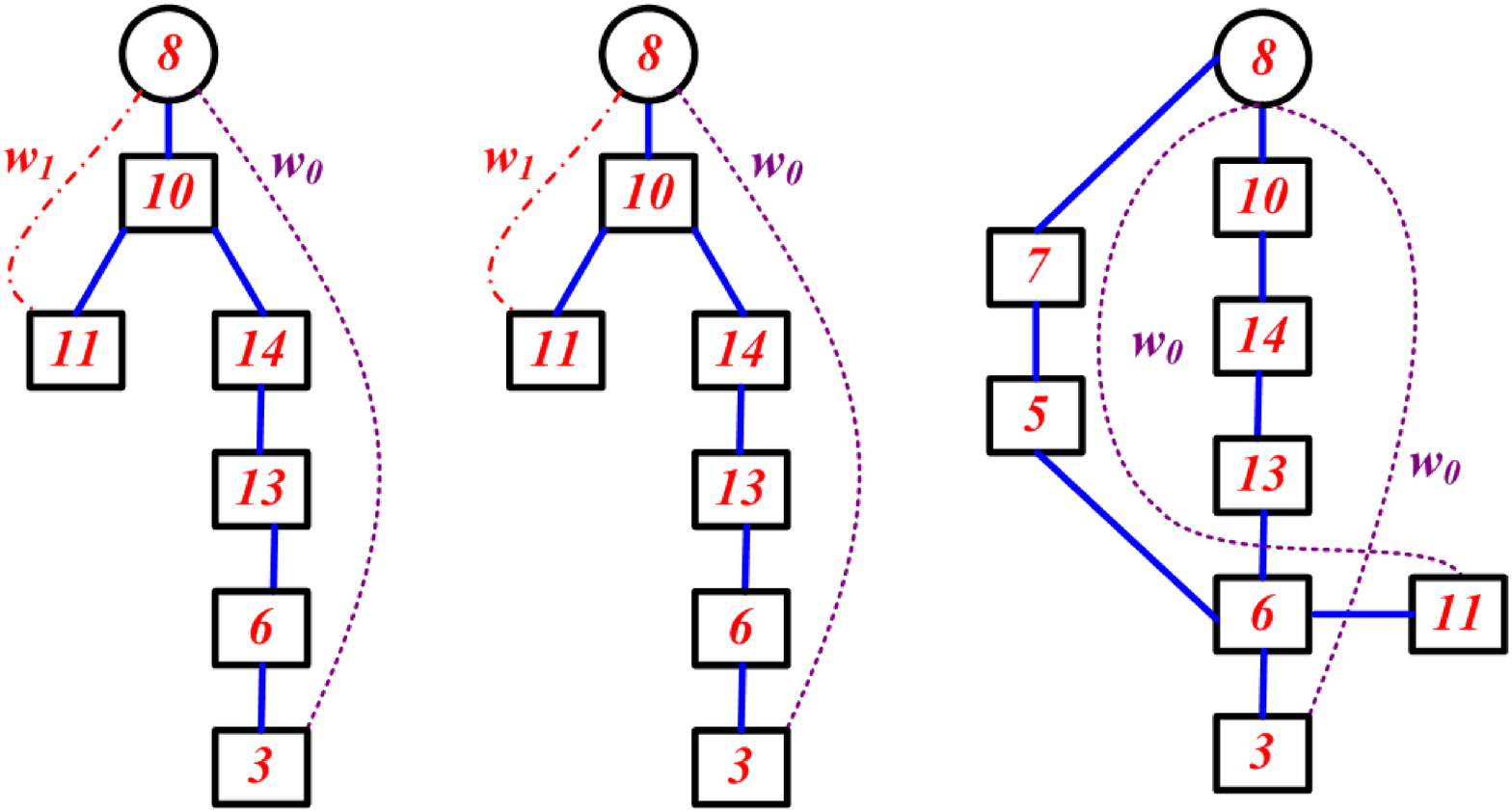}
        \\
        \mbox{(a)}\hspace{.9in}\mbox{(b)}\hspace{.9in}\mbox{(c)}
        \caption{For multicast session $ms_{3}$, (a) Light-trees built by Member-Only; (b) Light-trees built by Graph Renewal Strategy; (c) Light-hierarchy built by Extended Graph Renewal Strategy}
        \label{fig: hierarchy}
\end{figure}

Due to its TaC capability, an MI node is able to connect only one successor in a light-tree. However, for an MI node with high degree (at least of 4), two signals on the same wavelength from two different incoming ports can be switched to two different outgoing ports without any conflict (for instance in Fig.~\ref{fig: hierarchy}(c), two signals on the same wavelength $w_{0}$ from the source 8 traverse MI node 6 twice through two cross paths to reach destination 3 and 11). As a result an MI node could be visited twice in the light distribution structure by making use of different input and output port pairs. In this case, the multicast structure will be no longer a light-tree, but a light-hierarchy, where cycles may exist (c.f., Fig.~\ref{fig: hierarchy}(c)). A light-hierarchy is an extension of the lightpath, which is covered by only one wavelength. By benefiting from the particular capacity of 4-degree MI nodes, more destinations could be spanned by a light-hierarchy and fewer light-hierarchies will be required compared to light-trees. Hence the link stress could be improved. As fewer wavelengths a multicast session requires, more multicast sessions may be accepted in the network, which may lead to the improvement of network throughput also.

The light-hierarchy structure overcomes the inherent shortcoming of the light-tree structure, since a 4-degree MI node can be visited more than once in a light-hierarchy ($LH$). Nevertheless a link already in a sub $LH$ cannot be used any more on the same $LH$. In order to compute a light-hierarchy, Graph Renewal Strategy can be employed too. But, the topology renewal operation should be modified. At iteration $i$, only the edges in the shortest path newly added to $LH$ are delete from $G_{i}$, which then generates a new graph $G_{i+1}$ for the next iteration.

\textbf{Example 3:} Multicast session $ms_{3}\big(s:8,D:(3, 6, 10, 11, 13, 14)\big)$ is needed in the NSF network. Only the source is an MC node. Applying Member-Only algorithm~\cite{xjzhang2000}, node 10 is first added to the light-tree. Since both nodes 11 and 14 have the same distance of 1 hop to node 10, there are two possibilities. On one hand, if node 14 is connected to node 10 at first, then the light-trees in Fig.~\ref{fig: hierarchy}(a) may be obtained by Member-Only with the adding order of nodes: 8-10-14-13-6-3, 8-10-11. The same light-trees in Fig.~\ref{fig: hierarchy}(b) will be obtained by Graph Renewal Strategy too with the same adding order of nodes. This is because that node 10 is deleted from graph $G_{1}$ after node 14 connects to it, and 4-degree MI node 6 is deleted from graph $G_{4}$ after node 3 connects to it. At this moment, node 11 is an isolated node in the new graph $G_{5}$. Hence, it could not be spanned in the current light-tree and another light-tree should be built. However, with the help of light-hierarchy, the constraint could be relaxed. To generate a new graph, only the used edges are deleted from the previous graph. 4-degree MI node 6 is still retained in the new graph and so are the edges (6-11) and (6-5). It is easy to find the path $P(8-7-5-6-11)$ for node 11 in the new graph with Dijkstra's algorithm. So, the light-hierarchy in Fig.~\ref{fig: hierarchy}(c) benefits from the 4-degree MI node 6. It is able to save one wavelength. On the other hand, if node 11 is assumed to be connected to node 10 earlier than 14, Member-Only algorithm still needs two wavelengths while the others require only one.

\subsection{Proposed Algorithms}
\label{subsec: Proposed Algorithms}
Based on the above strategies, we propose two multicast routing algorithms with two different structures in WDM networks: Graph Renewal \& Distance Priority Light-Tree algorithm (GRDP-LT) and Extended Graph Renewal \& Distance Priority Light-Hierarchy algorithm (GRDP-LH). The difference between them is the strategy of graph generation operation (c.f., step-13 in Algorithm 1), which corresponds to different light-structures. In a light-hierarchy, the inherent shortcoming of the light-tree structure is overcome. That is why it is able to achieve the lowest link stress (c.f., Fig.~\ref{fig: link stress}).

 \begin{algorithm}[!t]
    \algsetup{indent=2em}
    \caption{Graph Renewal \& Distance Priority Light-Tree Algorithm (GRDP-LT) / (GRDP-LH)}
    \label{alg1}
    \begin{algorithmic}[1]
    \REQUIRE {A graph $G(V,E,c,W)$ and a multicast session $ms(s,D_{0})$.}
    \ENSURE {A set of Light-structures $LS_{k}$ each on a different wavelength $w_{k}$ for $ms(s,D_{0})$.}
    \STATE {$k \leftarrow 1$, $D \leftarrow D_{0}$}
    \WHILE {$(D\neq \O)$}
        \STATE {$i \leftarrow 1$} \COMMENT {$i$ is the serial number of a renewed graph}
        \STATE {$G_{i} \leftarrow G$, $MC\_SET \leftarrow \{s\}$, $LS_{k} \leftarrow \{s\}$}
        \WHILE {$(D$ is reachable from $MC\_SET$ of $LS_{k})$}
            \STATE {Find the nearest destination \textbf{$d_{i}$} to $LS_{k}$, and choose the optimal connector node \textbf{$c_{i}$} for \textbf{$d_{i}$}
            \begin{enumerate}
             \item Compute all the shortest path $SP_{G_{i}}(d,c)$ in $G_{i}$ from each $d \in D$ to $c \in MC\_SET$
             \item Find the nearest destination \textbf{$d_{i}$} to $LS_{k}$ such that
             \begin{equation}
             \label{eqation: sp-min}
                    c\big(SP_{G_{i}}(d_{i}, c)\big) = \min_{d \in D, c \in MC\_SET}{c\big(SP_{G_{i}}(d,c)\big)}
             \end{equation}
             \COMMENT {Function $c()$ is the cost of a path}
             \item Find the nearest connector node \textbf{$c_{i}$} to source $s$ in $LS_{k}$, if there are several connector nodes satisfying equation~(\ref{eqation: sp-min})
            \end{enumerate}
            }
            \STATE {Add $SP_{G_{i}}(d_{i}, c_{i})$ to $LS_{k}$}
            \STATE {$D \leftarrow D \setminus \{d\}$}
            \STATE {Add \textbf{$d_{i}$}  and MC nodes in $SP_{G_{i}}(d_{i}, c_{i})$ to $MC\_SET$}
            \IF {(\textbf{$c_{i}$} is an MI node)}
                 \STATE {Remove \textbf{$c_{i}$} from $MC\_SET$}
            \ENDIF
            \STATE {Generate a new graph $G_{i+1}$ from $G_{i}$.

            \begin{description}
              \item[GRDP-LT:]Delete all the non-leaf MI nodes and edges in $SP_{G_{i}}(d_{i},c_{i})$ from $G_{i}$, except $d$ if it is an MI node.
              \item[GRDP-LH:]Only delete the edges in $SP_{G_{i}}(d_{i},c_{i})$ from $G_{i}$.
            \end{description}
            }
            \STATE {$i \leftarrow i+1$}
        \ENDWHILE
        \STATE {Assign wavelength $w_{k}$ to $LS_{k}$}
        \STATE {$k \leftarrow k+1$} \COMMENT {Star a new light-structure $LS_{k+1}$}
    \ENDWHILE
    \end{algorithmic}
    \end{algorithm}

\section{Wavelength Assignment}
\label{sec: Wavelength Assignment}
The wavelength assignment problem (WAP~\cite{xhJia2001}) is always accompanied with the routing problems in WDM networks. It aims to assign wavelengths to a set of routes so that the number of wavelengths required can be minimized. Hence, the strategy for WAP also greatly impacts the performance of the routing algorithms. However, it is proved in~\cite{gWilfong1998} that WAP is NP-complete even in simple networks like rings or trees.

In our implementation, the First-Fit idea is employed. We search the wavelengths from index $1$ to $W$ (the maximum index), until we find the first wavelength index which is available on all the fiber links in a light-structure (i.e., light-tree or light-hierarchy). If and only if all the light-structures for a multicast session are assigned with a free wavelength index, this session could be accepted. Otherwise (i.e. no such wavelength index could be found), the multicast session will be blocked.

\section{Performance Evaluation And Simulation}
\label{sec: Performance Evaluation And Simulation}
\subsection{Simulation Model}
\label{subsec: Simulation Model}
From previous 3 examples, we can see the proposed algorithms work well in the NSF network. To show its flexibility, other topologies like USA Longhaul network (28 nodes, 7 nodes 4-degree and 1 node 5-degree) and European Cost-239 network (11 nodes, 4 nodes 4-degree, 6 nodes 5-degree and 1 node 6-degree) are employed as platforms for the simulations. In these topologies, without loss of generality each edge is associated with an equal cost of $1~unit~ hop-count~cost$ and an equal delay of $1~unit~ hop-count~delay$. For each fiber between two neighbor nodes, the number of wavelengths supported is denoted by $W$. It is set to $W = 20$ for the sake of short simulation time. The members of a multicast group and the MC nodes are assumed to be uniformly distributed in the topology. When simulating the performance of throughput of network, the multicast group size (include the source) is generated by a random variable following a uniform distribution in the internal $[3, N-1]$, where $N$ is the number of nodes in the network.

\subsection{Evaluation Metrics}
\label{subsec: Evaluation Metrics}
The following five metrics are considered.
\begin{itemize}
  \item \emph{Link Stress}. It is defined as the maximum number of wavelengths required per fiber link. For the case of a single multicast session, it equals to the number of light structures built.

  \item \emph{Average Delay}.  It is the average of end-to-end delays from all destinations to the source in a multicast session.


  \item \emph{Diameter}. It is defined as the maximum hop counts in the lightpath from each destination to the source in the light structures. For a light-tree, the diameter can be the maximum distance from the destinations and the source; while for a light-hierarchy, it is the length of the longest lightpath from all the destinations to the source. In a light-hierarchy, the diameter may be bigger than the maximum distance between the source and the destinations, because there may be cycles.


  \item \emph{Total Cost}. It is used to measure the wavelength channel cost consumed in order to establish a multicast session.


  \item \emph{Throughput}. It is computed as the maximum number of multicast sessions could be accepted, given the number of wavelengths supported per fiber. Here, it is set to $W = 20$.
\end{itemize}

\subsection{Performance Analysis}
\label{subsec: Performance Analysis}
In the first step of our simulation, the GRDP-LT algorithm is compared with the famous Member-Only algorithm (MO). Since both of these two algorithms produce light-trees for a multicast request, this step aims to show the performance improvement by using the proposed Graph Renewal \& Distance Priority algorithm. Then, the comparison of performance is done between two different multicast structures: light-tree and light-hierarchy (using GRDP-LH algorithm). From the comparison, we will verify whether light-tree structure is still optimal in sparse splitting WDM mesh network and evaluate the quality of light-hierarchy.

\subsubsection{MO versus GRDP-LT}
\label{subsubsec: MO versus GRDP-LT}

\begin{figure*}
    \begin{center}
    $\begin{array}{ccc}
    \epsfxsize=2.17in \epsffile{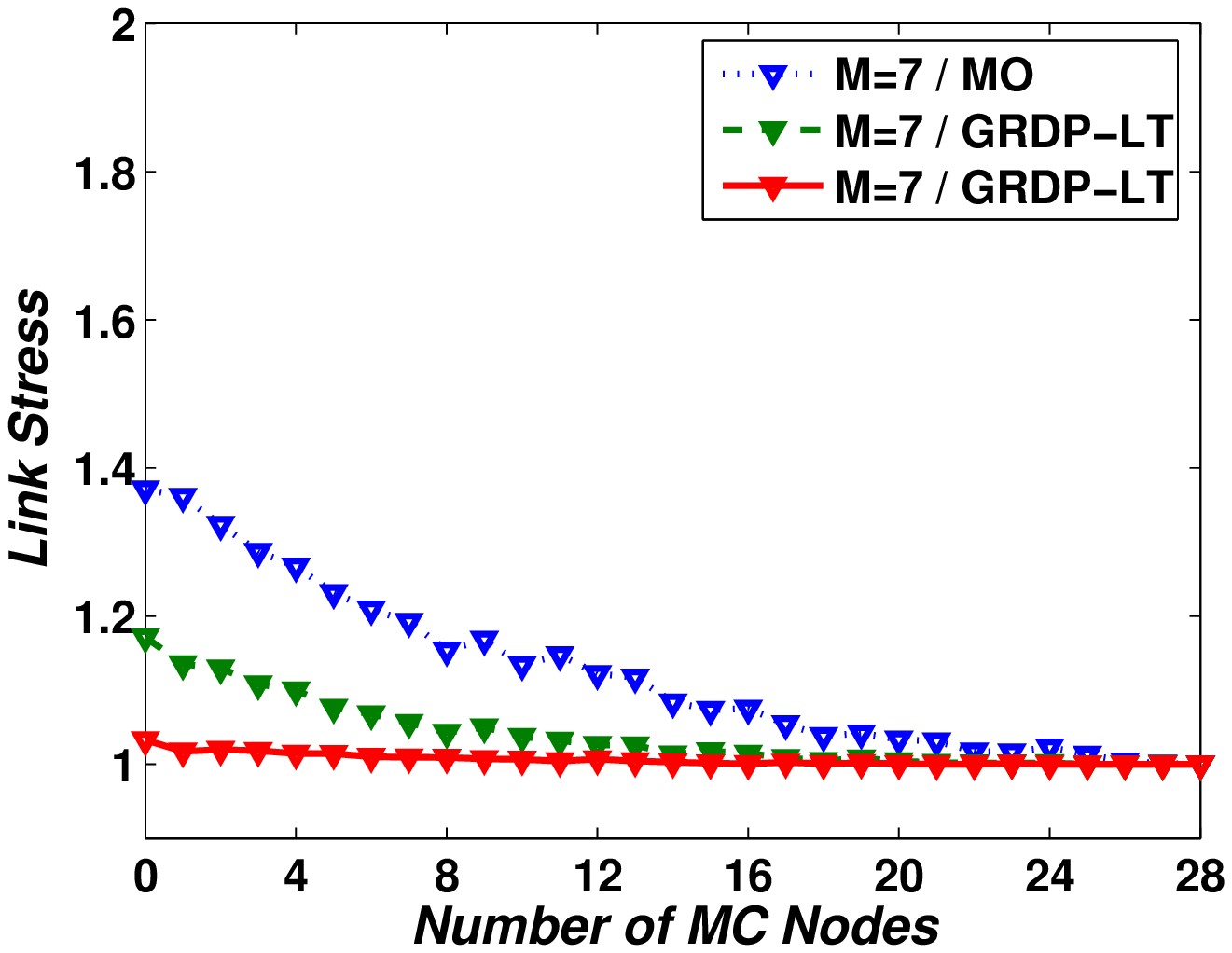}
  & \epsfxsize=2.17in \epsffile{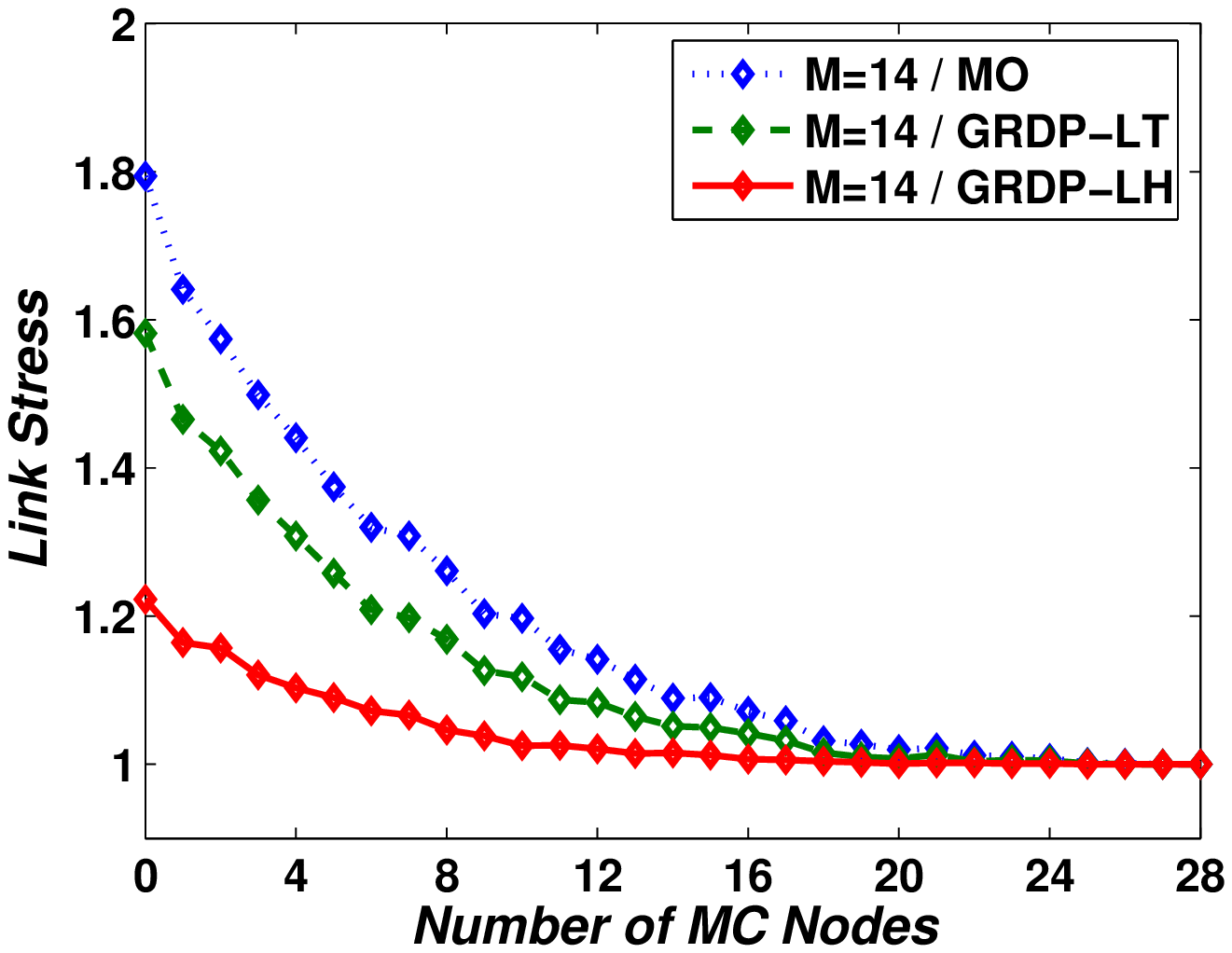}
  & \epsfxsize=2.17in \epsffile{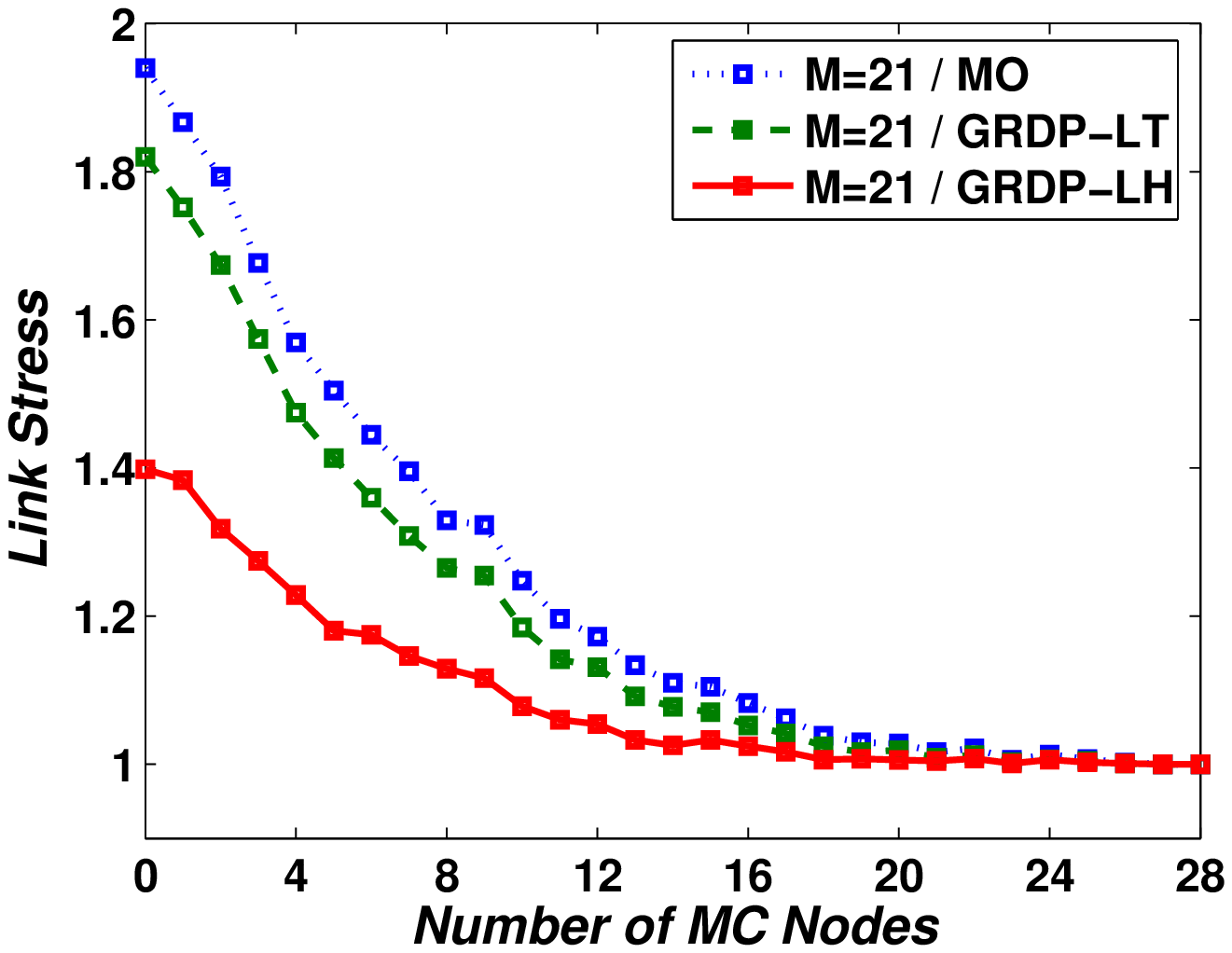} \\
    \mbox{\bf (a)} & \mbox{\bf (b)} & \mbox{\bf (c)}
    \end{array}$
    \end{center}
    \caption{Comparison of Link Stress in the USA Longhaul topology when multicast member (a) ratio = 25\%; (b) ratio = 50\%; (c) ratio = 75\%.}
    \label{fig: link stress}
    \end{figure*}

\begin{figure*}
    \begin{center}
    $\begin{array}{ccc}
    \epsfxsize=2.17in \epsffile{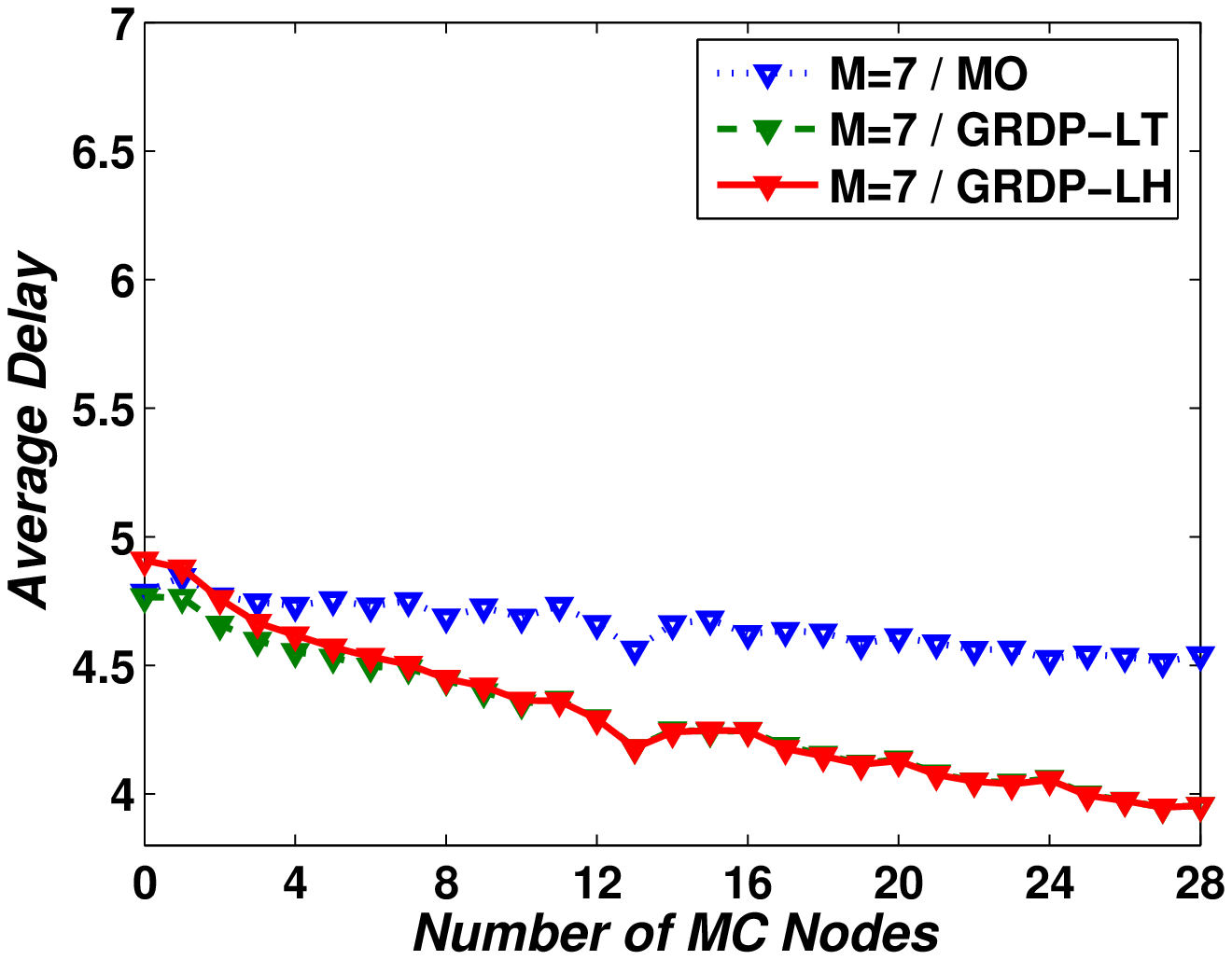}
  & \epsfxsize=2.17in \epsffile{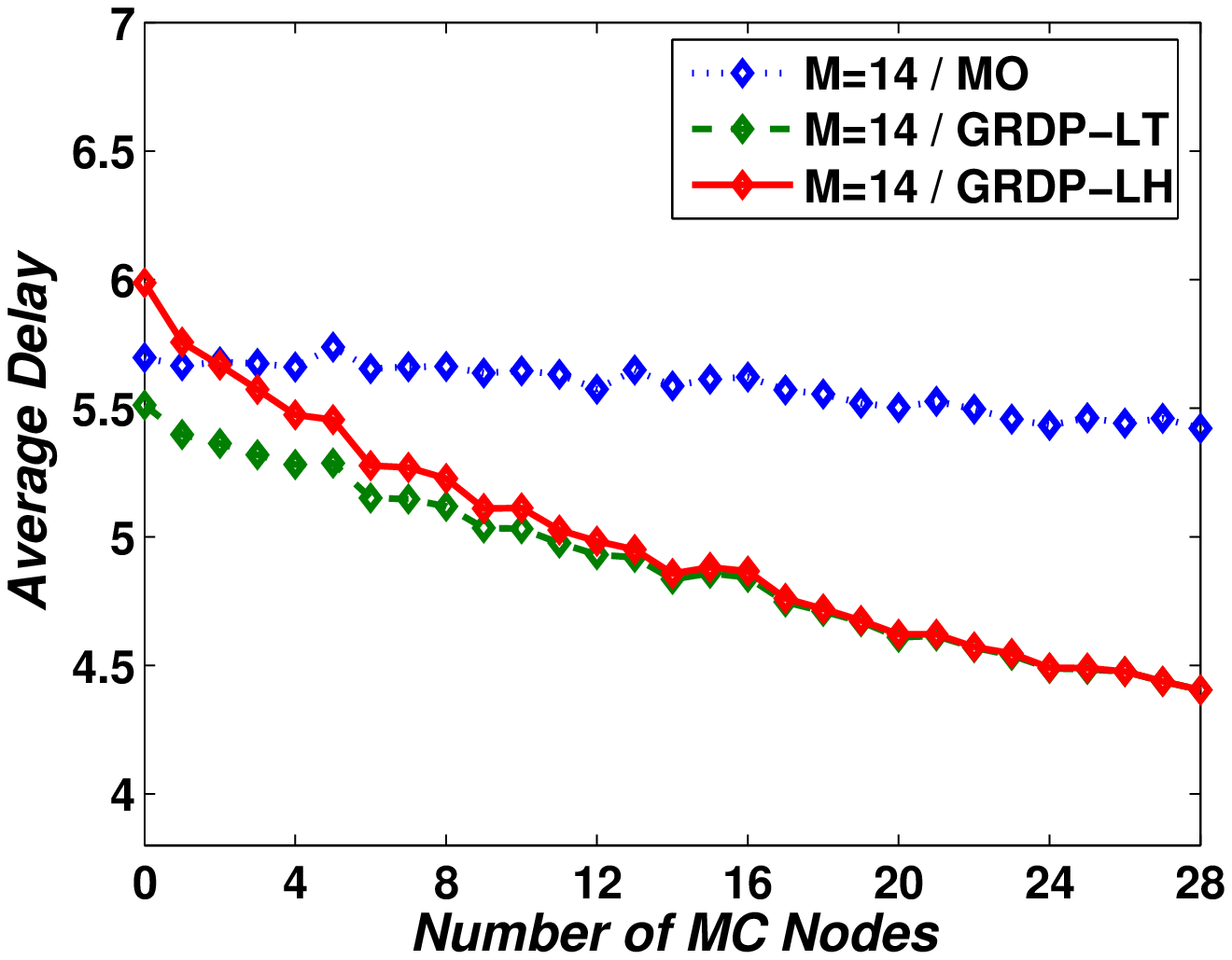}
  & \epsfxsize=2.17in \epsffile{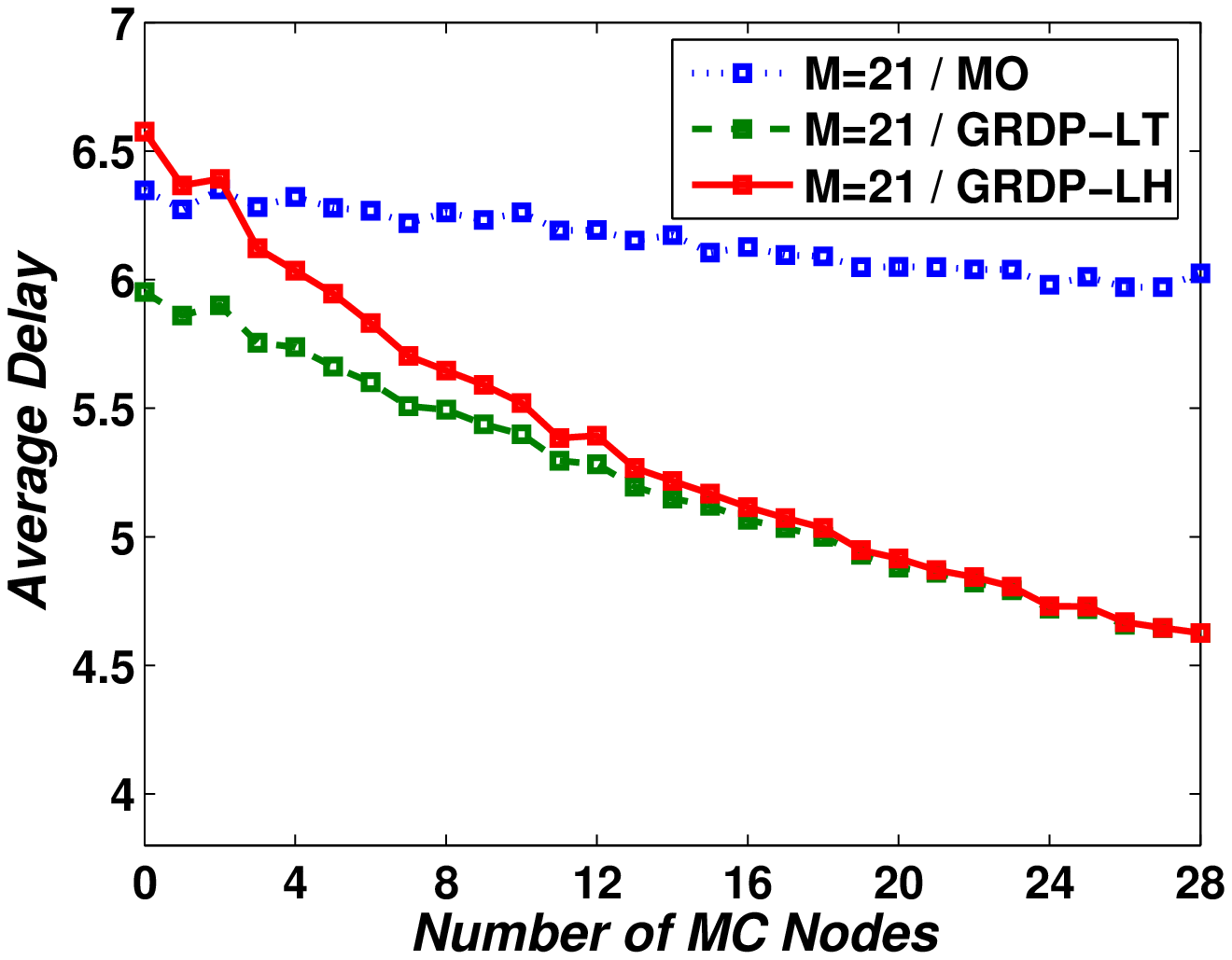} \\
    \mbox{\bf (a)} & \mbox{\bf (b)} & \mbox{\bf (c)}
    \end{array}$
    \end{center}
    \caption{Comparison of Average Delay in the USA Longhaul topology when multicast member (a) ratio = 25\%; (b) ratio = 50\%; (c) ratio = 75\%.}
    \label{fig: average delay}
    \end{figure*}

\begin{figure*}
    \begin{center}
    $\begin{array}{ccc}
    \epsfxsize=2.17in \epsffile{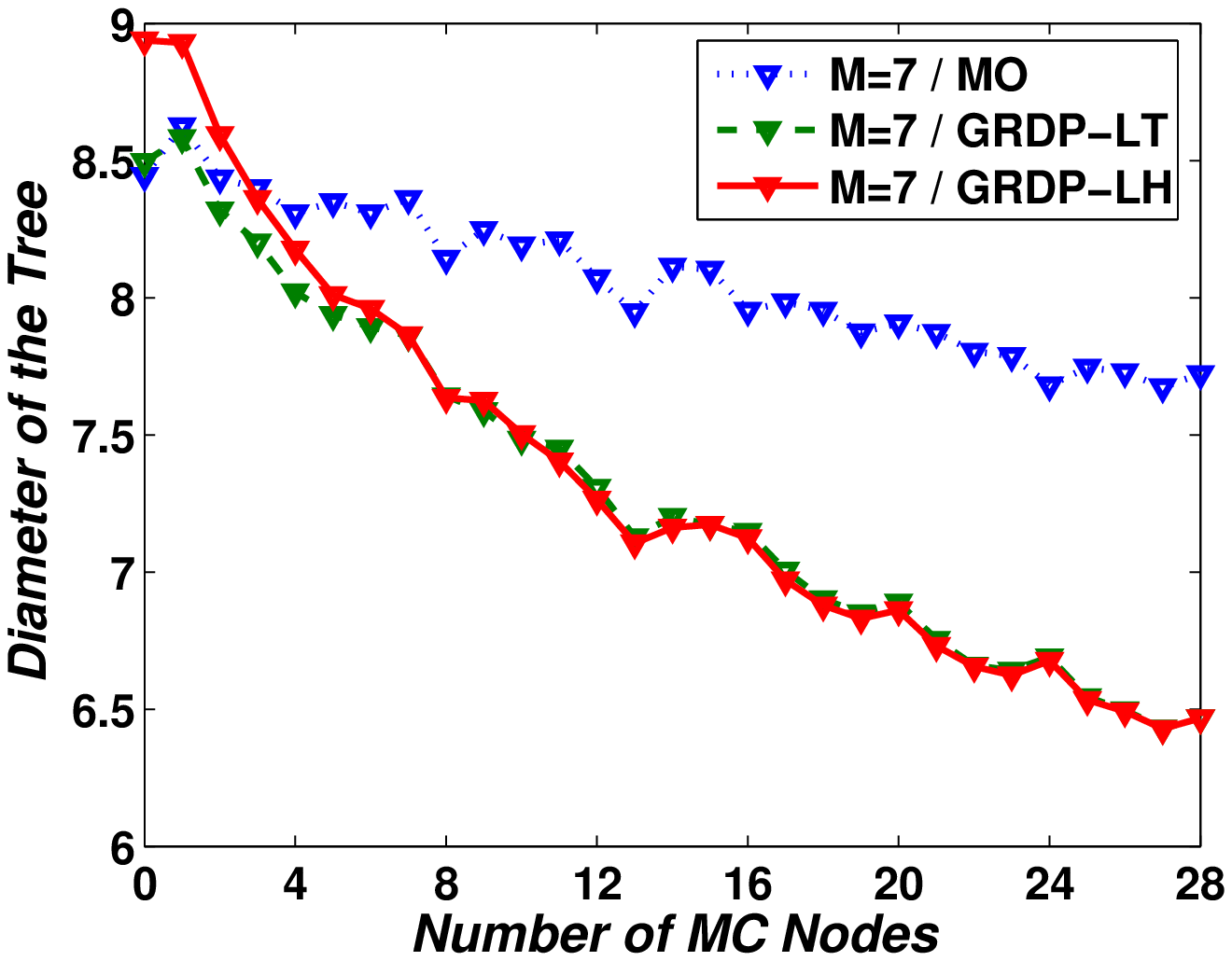}
  & \epsfxsize=2.17in \epsffile{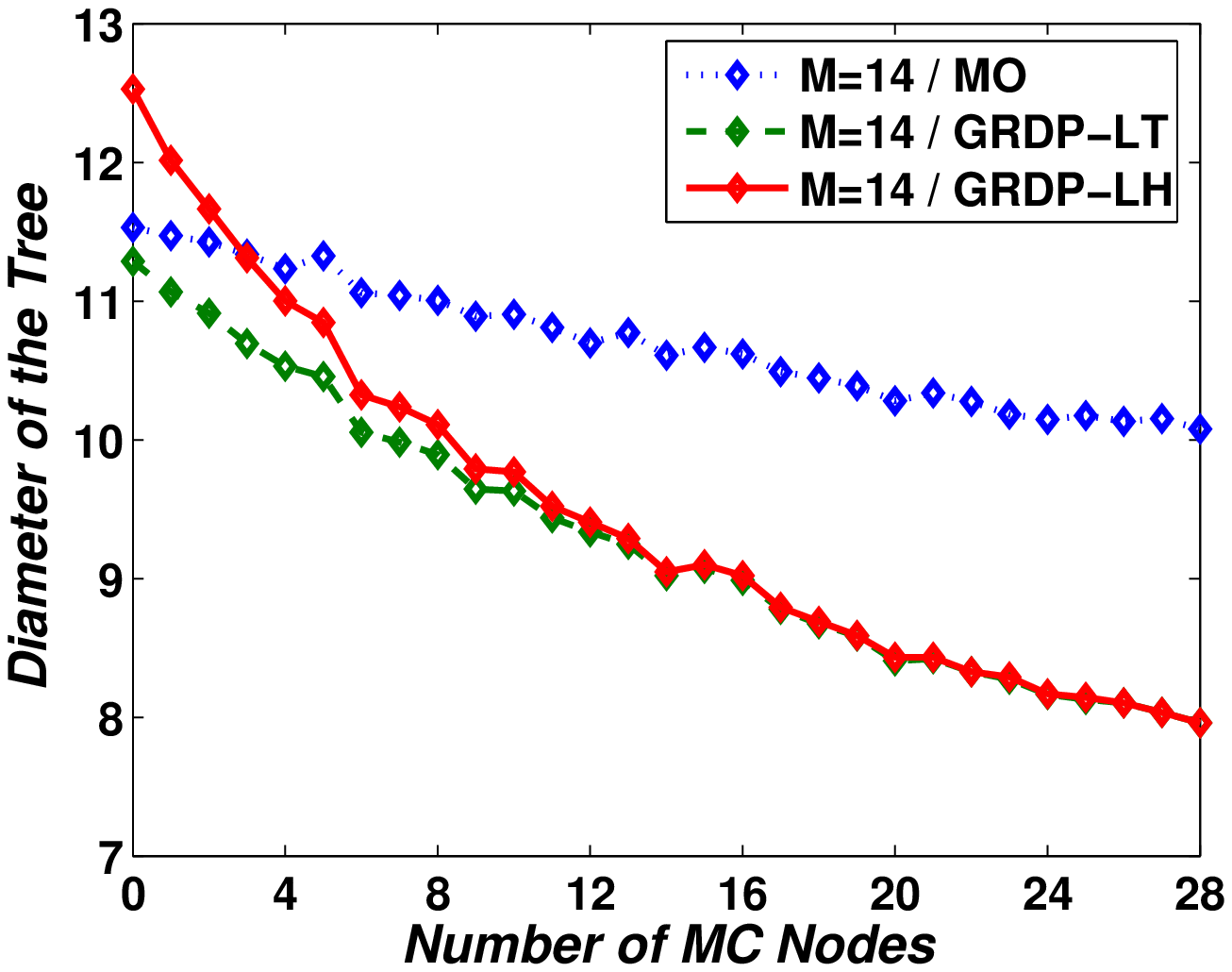}
  & \epsfxsize=2.17in \epsffile{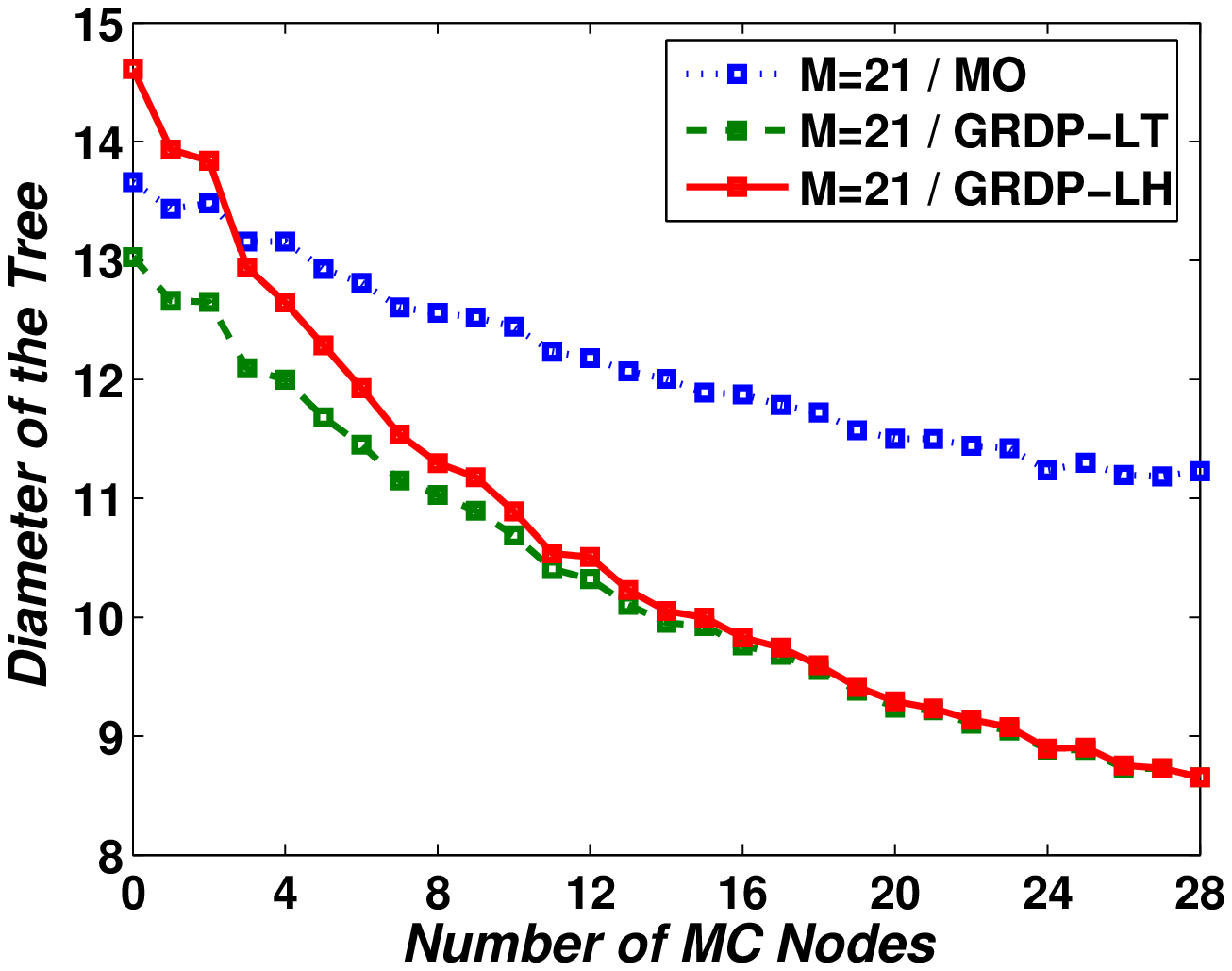} \\
    \mbox{\bf (a)} & \mbox{\bf (b)} & \mbox{\bf (c)}
    \end{array}$
    \end{center}
    \caption{Comparison of Diameter in the USA Longhaul topology when multicast member (a) ratio = 25\%; (b) ratio = 50\%; (c) ratio = 75\%.}
    \label{fig: diameter}
    \end{figure*}

In Figs.~\ref{fig: link stress}-\ref{fig: throughput}, the results of simulations in the USA Longhaul topology and the European Cost-239 topology are presented.

(i) As plot in Fig.~\ref{fig: link stress}, the link stress of GRDP-LT light-tress is always lower than MO, reduced up to 15\%, 12\% and 6\% (calculated by (MO-GRDP)/MO) when the group size ($M$, count the source) equals to 7, 14 and 21 respectively. The reason can be explained as follows: since more available paths could be found to connect a destination to a light-tree, more destinations could be spanned in a light-tree and thus fewer light-trees are required for a multicast session.

(ii) In Figs.~\ref{fig: average delay} and \ref{fig: diameter}, both the average delay and the diameter of light-trees for GRDP-LT are smaller than MO. Furthermore, it is not difficult to find that the reduction of the average delay and the diameter become significant (up to 13\%, 19\% and 23\% for average delay respectively when M=7, 14 and 21; and up to 16\%, 21\% and 23\% for diameter of light-trees respectively when M=7, 14 and 21), when the number of MC increases. It is because that the In Tree Distance Priority operative only when there are enough MC connector nodes for a chosen destination to join the currently multicast light-tree. And, the preconditions to produce enough choices of connector MC nodes are: first the proportion of MC nodes in the network is high enough, and second there are sufficient destinations in a multicast session.

(iii) As shown in Fig.~\ref{fig: throughput}(a), the total cost of GRDP-LT is slightly better than MO in any situation. This is because both these two algorithms apply the Minimum Path Heuristic~\cite{hTakahashi1980}.
(iv) From the point of view of the throughput, GRDP-LT is able to stand a little more multicast sessions simultaneously than MO as shown in Figs.~\ref{fig: throughput}(b)(c).

\subsubsection{Light-tree versus Light-hierarchy}
\label{subsubsec: Light-tree versus Light-hierarchy}
~\\
(i) As plotted in Fig.~\ref{fig: link stress}, if there is no MC node in the network, the link stress of light-hierarchies is 0.14, 0.36 and 0.42 respectively smaller than GRDP-LT light-trees when M=7, 14 and 21. It is very interesting to find that the light-hierarchy structure is able to reduce the link stress more and more as the number of members grows. The advantage of light-hierarchies is even more evident in the sparse light splitting case.

(ii) We can also see in the Figs.~\ref{fig: average delay} and \ref{fig: diameter} before the number of MC nodes grows larger than 3 (corresponds to 10\% of MC nodes), the average delay and the diameter of light-hierarchies is bigger than GRDP-LT, and even than MO. Fortunately, when the number of MC nodes is above 4, these two parameters for light-hierarchies decrease to below MO, and also approach to GRDP-LT until they reach the same value. The reason is that, when there is no MC node or the MC nodes are too sparse, in order to include more destinations in one light-hierarchy and thus to reduce the link stress, longer paths should be employed to connect destinations which cannot be connected by using the shortest path as done in Member-Only algorithm. And, in case that the proportion of MC nodes is high enough, the In Tree Distance Priority works well.

\begin{figure*}
    \begin{center}
    $\begin{array}{ccc}
    \epsfxsize=2.17in \epsffile{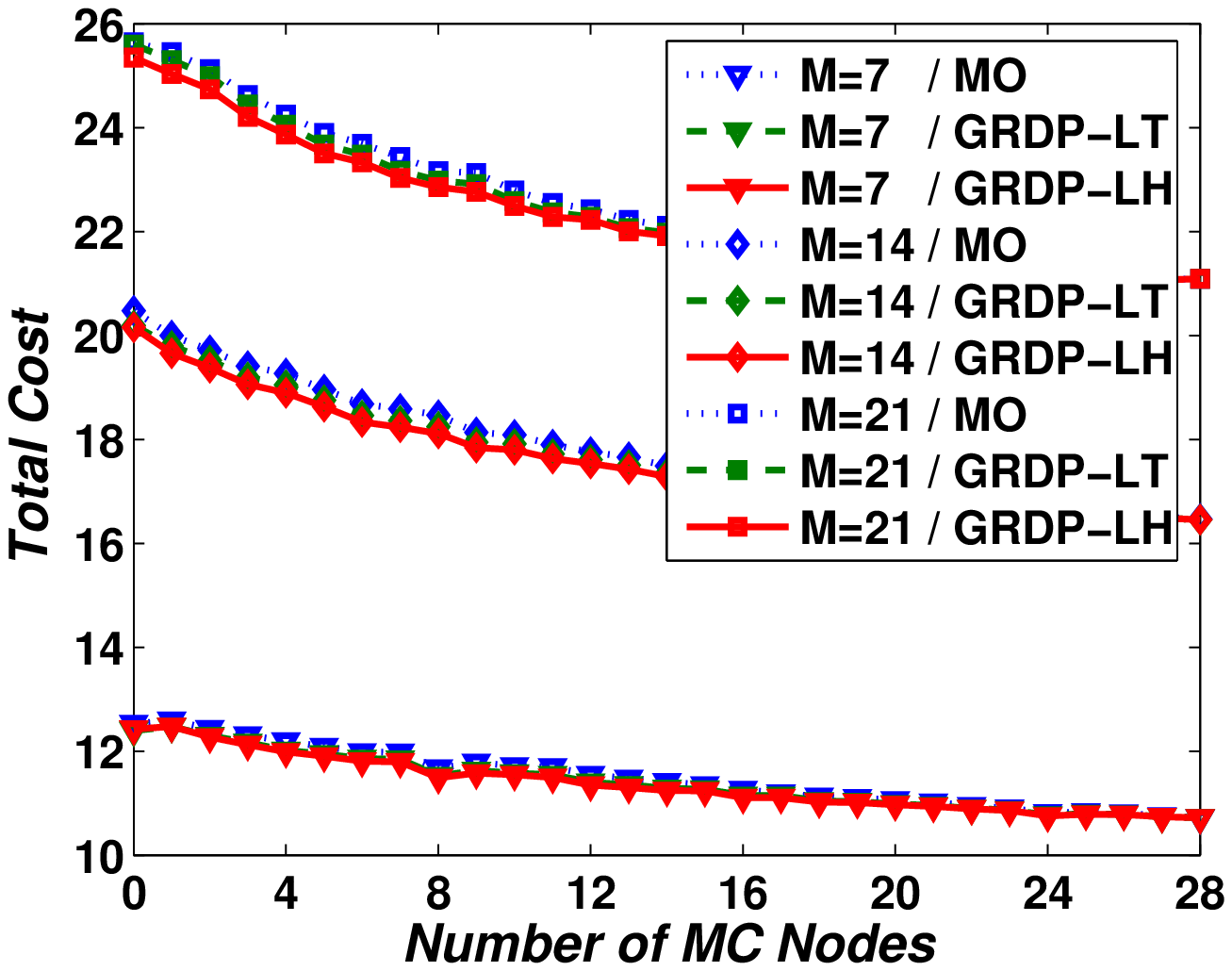}
  & \epsfxsize=2.17in \epsffile{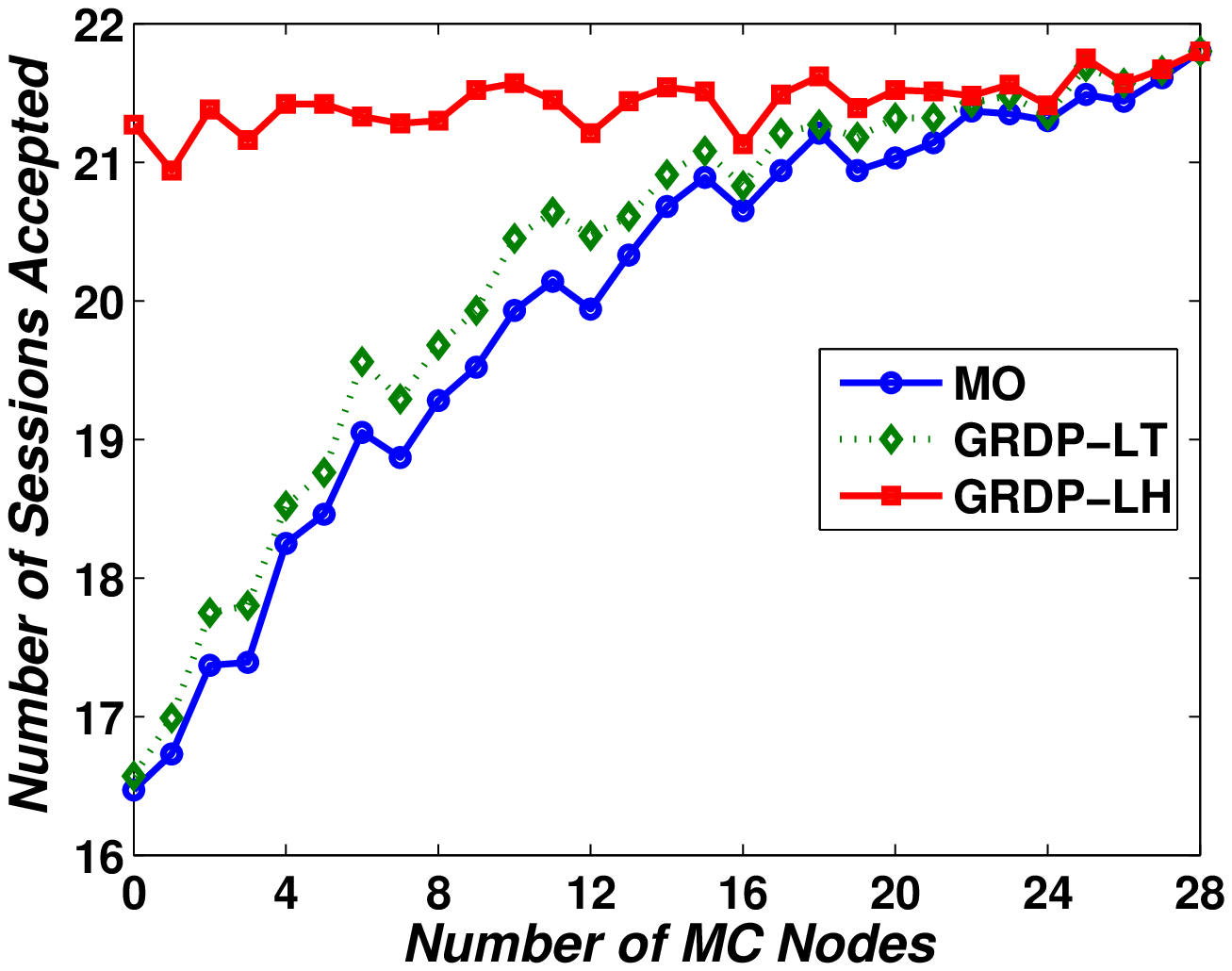}
  & \epsfxsize=2.17in \epsffile{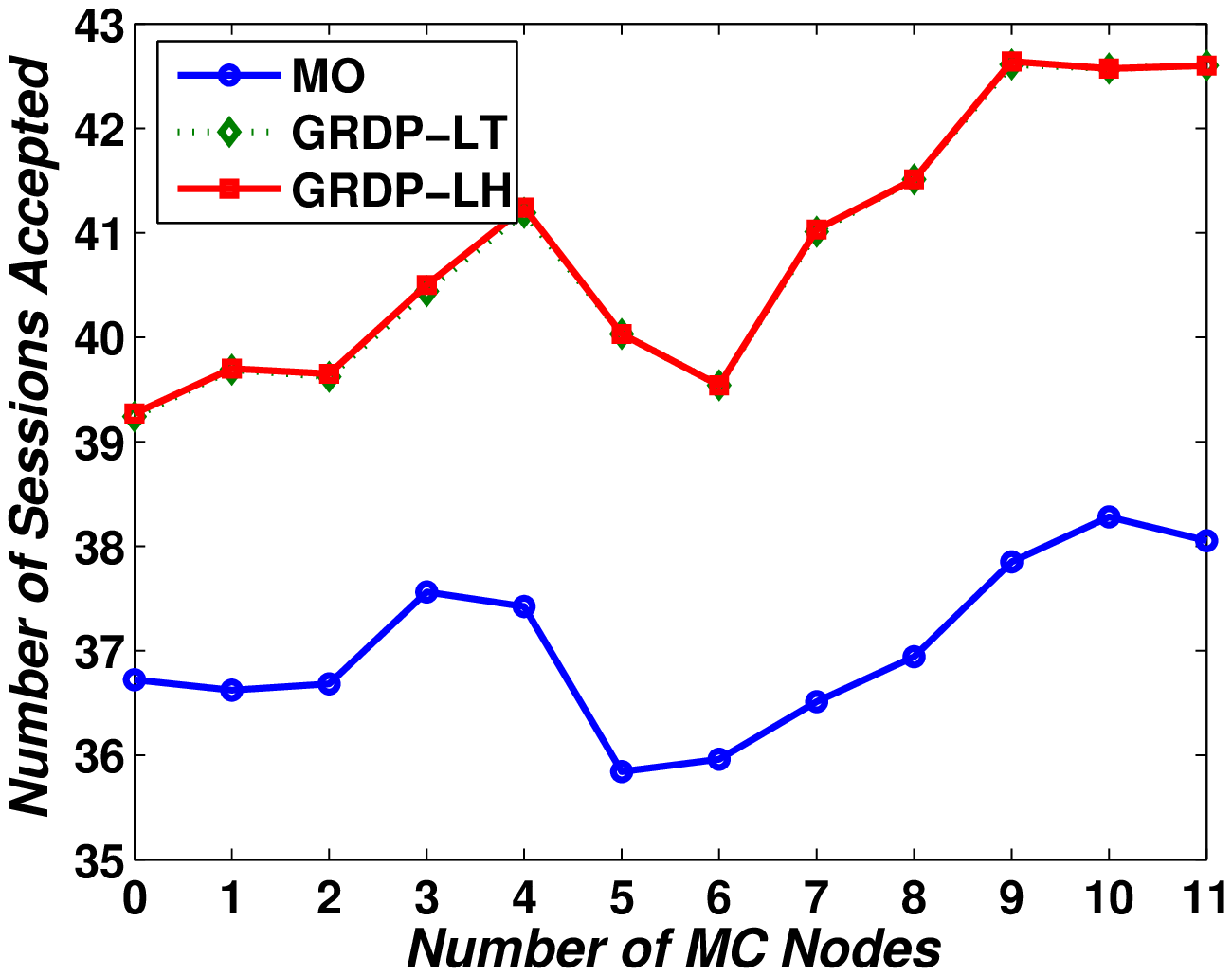} \\
    \mbox{\bf (a)} & \mbox{\bf (b)} & \mbox{\bf (c)}
    \end{array}$
    \end{center}
    \caption{(a) Comparison of Total Cost in the USA Longhaul topology; (b) Comparison of Throughput in the USA Longhaul topology; (c) Comparison of Throughput in the Cost-239 topology.}
    \label{fig: throughput}
    \end{figure*}

(iii) As far as the total cost indicated in Fig.~\ref{fig: throughput}(a), light-hierarchy structure achieves almost the same or slightly better than GRDP-LT, not even to say than MO.

(iv) Regarding throughput, up to 4.7 additional multicast sessions (an improvement of 22\%) can be accepted by the light-hierarchy structure compared to GRDP-LT as plotted in Fig.~\ref{fig: throughput}(b). And whatever the number of MC nodes is, the light-hierarchy can accommodate more additional multicast sessions than both GRDP-LT and MO. Moreover, in order to study the throughput versus the number of 4-degree MI nodes in the topology, we also plot the number of multicast sessions accepted before blocking in European Cost-239 network, where all 11 nodes have a degree of at least 4. As shown in Fig.~\ref{fig: throughput}(c), the light-hierarchy structure has accepted the same number (39.5, when 50\% of nodes are MC) of multicast sessions as GRDP-LT. European Cost-239 is a network with high connectivity, generally only one light-tree is generally enough to accommodate all multicast members with GRDP-LT algorithm. Hence, it is reasonable that GRDP-LH has the same performance as GRDP-LT in term of throughput when all network nodes have 4 degree or above.

\subsubsection{Is Light-tree Structure Optimal?}
\label{subsubsec: Is Light-tree Structure Optimal?}
From the two comparisons above, we can see that although the Graph Renewal strategy could be used to improve the quality of light-trees, the improvement is limited. This limitation is mainly due to the inherent drawback of light-tree structure. With help of the light-hierarchy structure, the constraint is relaxed to delete used links. By benefiting from the at least 4-degree MI nodes, a light-hierarchy has an even bigger capacity to span more destinations. Thus link stress and network throughput could be greatly improved again. Based on the analysis and the numeric results, it is obvious that the light-tree structure is no longer optimal in terms of link stress and throughput, but the proposed light-hierarchy structure can be better.

\section{Conclusion}
\label{sec: Conclusion}
In this paper, we examined the multicast routing and wavelength assignment problem in all-optical WDM networks with sparse light splitting. The Graph Renewal Strategy is first introduced to improve the quality of light-trees, which deletes the constraint nodes from the network topology. By spanning the nearest destination with the shortest path in the renewed graph, the Graph Renewal Strategy diminishes the link stress and total cost. It also gains a higher network throughput than the currently most efficient algorithm. Then, the In Tree Distance Priority is applied to reduce the average delay and the diameters of light-trees. However, the improvement of the Graph Renewal Strategy is limited due to the inherent drawback of the light-tree structure. Thereby, a new multicast structure called light-hierarchy is proposed.
A light-hierarchy is an extension of a light-tree, while it accepts cycles. With the help of the light-hierarchy structure, the constraint of multicasting is relaxed, and accordingly the Graph Renewal Strategy is extended to compute light-hierarchies. Simulations showed that the performance in terms of link stress and network throughput is greatly improved again by light-hierarchies, while consuming the same wavelength channel cost. Therefore, the light-tree structure is not optimal, but the light-hierarchy structure can be better for multicast routing in sparse light splitting WDM networks.


%
%



%

\end{document}